# Origin of Perpendicular Magnetic Anisotropy in Yttrium Iron Garnet Thin Films Grown on Si (100)

Zurbiye Capku,[1,2] Caner Deger,[3] Perihan Aksu,[4] Fikret Yildiz [1]

[1]Department of Physics, Gebze Technical University, Gebze, Kocaeli, 41400, Turkey
[2]Department of Physics, Boğaziçi University, Beşiktaş, Istanbul, 34342, Turkey
[3]Department of Physics, Marmara University, Kadikoy, Istanbul, 34722, Turkey
[4]Institute of Nanotechnology, Gebze Technical University, Gebze, Kocaeli, 41400, Turkey

We report the magnetic properties of yttrium iron garnet (YIG) thin films grown by pulsed laser deposition technique. The films were deposited on Si (100) substrates in the range of 15-50 nm thickness. Magnetic characterizations were investigated by ferromagnetic resonance spectra. Perpendicular magnetic easy axis was achieved up to 50 nm thickness. We observed that the perpendicular anisotropy values decreased by increasing the film thickness. The origin of the perpendicular magnetic anisotropy (PMA) was attributed to the texture and the lattice distortion in the YIG thin films. We anticipate that perpendicularly magnetized YIG thin films on Si substrates pave the way for a cheaper and compatible fabrication process.

## I. INTRODUCTION

Magnetic garnet films have recently begun to take the place of conducting ferromagnetic materials in spintronic applications. The insulating features of the garnets eliminates the disadvantages of Eddy currents, which causes loss of information in relevant applications [1]. They have attracted great attention for high frequency and fast switching of magnetic properties [2]. In particular, perpendicular magnetization in garnet films is very crucial for the field of spintronics i.e. spin-orbit switching, spin transfer torque and, a reliable and rapid response [3, 4]. Yttrium iron garnet (YIG) is considered to be one of the most important magnetic insulators. Static and dynamic magnetic properties of bulk crystal or YIG films in the micrometer thickness range have been investigated in great detail and widely used in microwave applications (filtering, tunabling, isolators, phase shifters, etc.) [5, 6]. However, the process of thin/ultrathin YIG films plays a key role for spintronic [7-11] and magneto-optical applications [12-14].

Many spintronic applications require a fine tuning of the orientation and magnitude of the magnetic anisotropy [15, 16]. Perpendicular magnetic anisotropy (PMA) has led to a revolutionary breakthrough in the technology such as the invention of high-density Magnetoresistive Random-Access Memory devices (MRAM). The effective control over the magnetic anisotropy leads to highly remarkable features such as increased data storage capacity in the magnetic recording media, magnon transistor [17], and advancement in the logic devices [16]. Despite the fact that enhancement of PMA in metal thin films is a well-established phenomenon [18, 19], generating PMA in insulating materials such as YIG remains a challenge. For such reasons, ferromagnetic insulators with PMA have been of particular importance for both fundamental scientific research and technological applications. Recent developments in the magnonic field have attracted great attention to ultrathin/thin YIG films perpendicularly magnetized. For example; YIG with the possess of PMA, has a unique feature in spin-orbit torque (SOT) applications [20]. The typical anisotropy in YIG films is in-plane anisotropy (IPA) which mainly originates from the strong shape anisotropy. When the magnetocrystalline anisotropy overcomes the shape anisotropy, the direction of the magnetic easy axis switches to the out of the film plane, resulting in PMA. In the literature, the control on magnetic anisotropy in YIG thin films has been studied by various substrate, temperature, and thicknesses [21, 22]. PMA in YIG films were achieved by using a buffer layer [23] and/or doping with rare earth elements [1, 24, 25]. In these studies, garnet substrates such as Yttrium aluminium garnet (YAG) [26] and Gadolinium gallium garnet (GGG) were used to grow epitaxial YIG thin films due to their similar crystalline structure [22, 27]. Lattice constants of YIG film and GGG substrate are $a_{YIG}$= 12.376 Å and $a_{GGG}$= 12.383 Å, respectively [28]. This lattice match between YIG and GGG provides high quality crystallized YIG films [29]. However, the use of the GGG substrate in certain areas is limited and also costs much for large area applications. The use of Silicon (Si) as a substrate has many advantages; cost-effectiveness and widespread use in electronic devices and integrated circuits. Si has an fcc diamond cubic crystal structure with a lattice constant of 5.43 Å. The nearest neighbor distance between two Si atoms is 2.35 Å [30]. On the other hand, YIG has a cubic structure consisting of $Y^{3+}$ ions in dodecahedral (c) sites, $Fe^{3+}$ ions in tetrahedral (d) and octahedral (a) sites in polyhedron of oxygen ions [31]. The nearest interionic distance in YIG is reported as ($Y^{3+}$ - $O^{2-}$) at 2.37 Å [31]. The atomic distances are comparable; thus, one can achieve crystalline / texture YIG on Si (100).

In this study, we report the PMA enhancement in YIG films grown on Si substrates by pulsed laser deposition (PLD) technique. Several parameters such as oxygen pressure, substrate temperature, post-annealing treatment, and laser power play an important role in the stoichiometry and crystallinity of the YIG films fabricated by PLD. In this report, YIG films with different thicknesses were grown on Si (100) substrates. A post-annealing process was carried out for all films to improve the crystallization and substrate-film lattice mismatch. The effect of the thickness on the magnetic anisotropy values was studied. In some reports, PMA in YIG films was obtained in the thickness range of 10-20 nm [22, 23, 32]. However, in this study, the lattice distortion/texture in YIG films gave rise to PMA in 15-50 nm thickness. We anticipate that our study not only offers a basis for fundamental understanding but also will inspire the integration of perpendicularly magnetized YIG thin films with technological applications.

## II. EXPERIMENTAL STUDIES

Before the thin film deposition, the oxide layer was first etched from the surface of Si (100) substrates with diluted Hydrofluoric (HF) acid for a few minutes. The substrates were further cleaned in acetone, methanol and Isopropyl alcohol for 15 minutes by using an ultrasonic bath. Subsequently, the surfaces were spray dried with Nitrogen gas. Following the chemical cleaning, the substrates were introduced into high vacuum chamber and annealed at 500 ºC for an hour. PLD with a KrF excimer laser, a Coherent COMPex Pro 205F operating at $\lambda$ = 248 nm (20 ns pulse duration) was used to obtain the desired YIG film stoichiometry by adjusting the oxygen pressure and deposition temperature. The base pressure of the deposition chamber was $1.0 \times 10^{-9}$ mbar. The commercial polycrystalline sintered YIG was used as the deposition target. The distance between the target and substrate was about 60 mm. The films were fabricated using laser energy of 220 mJ at a pulse repetition rate of 10 Hz in an oxygen atmosphere of $1.0 \times 10^{-5}$ mbar. The substrate temperature was 400 °C during growth. The deposition rate was 0.96 nm/min. The films were cooled within a rate of 9.6 ºC/min inside the chamber. Thereafter, the films were annealed at 850 °C for 2 hours in an air atmosphere and cooled down to room temperature by a ratio of 1.2 ºC/min. The thicknesses of the annealed films were defined as 15 nm, 20 nm, 35 nm and 50 nm using X-ray reflectivity (XRR) method.

## III. RESULTS

Atomic Force Microscopy (AFM) was performed for the surface morphology and roughness of the films. A representative AFM image of the annealed YIG film at 850 ºC with a root mean square (RMS) roughness value of 0.8 nm was given at the inset of Fig.1. Structural properties of the films were characterized by X-ray Diffraction (XRD) measurement using a Rigaku 2000 DMAX diffractometer with a Cu (alpha) wavelength of 1.54 nm. The θ-2θ scan XRD pattern was demonstrated in Fig.1. A typical (420) peak of YIG was observed for the annealed films. At each measurement, a signal from the sample plate was detected at 44º. The additional peaks around the (400) plane of the Si substrate correspond to the Kβ, Lα1, Lα2, Kα1 and Kα2 lines of the incident x-ray.

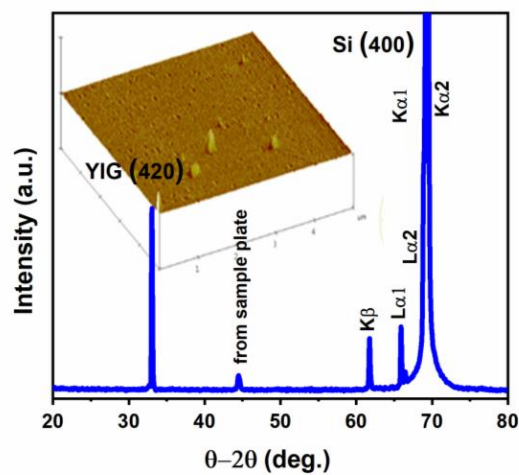

Fig. 1. XRD pattern of YIG thin film on a Si substrate. θ-2θ scan which shows the (420) characteristic peak of 20 nm YIG. (Inset: AFM image of a 20 nm YIG film.)

The chemical analysis was performed by X-ray photoelectron spectroscopy (XPS) measurement. The survey scan XPS spectrum is represented in Fig. 2(a). The spectrum confirmed the presence of Y, Fe and O elements on the surface of YIG film grown on Si. The fittings of the spectral ranges related to the elements which are used to determine the composition ratio are given in Figs. 2(b)-2(d). XPS spectrum of the Fe 2p region in Fig. 2(c) shows the valance state of the Fe ions. Y/Fe compositional ratio was found to be 0.59 and Fe/O ratio was 0.44. These values are very close to the bulk YIG compositional ratios within the experimental error (Y/Fe = 0.6 and Fe/O = 0.42 in bulk YIG) [11].

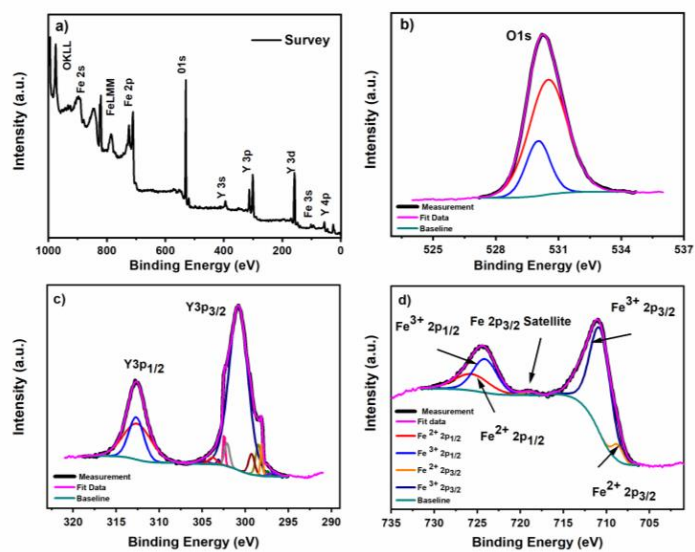

Fig. 2. XPS spectrum of YIG thin film grown on Si substrate. (a) XPS survey scan. Fitted XPS spectrum of (b) O 1s. (c) Y 3p. (d) Fe 2p.

Ferromagnetic resonance (FMR) measurements were performed to the annealed YIG films by an X-Band (9.1 GHz) JEOL series ESR spectrometer at room temperature. FMR is a powerful technique, which the analysis of the spectra also provides the values of anisotropy constants [33-35]. The resonance profile is determined by the field (H) derivative of the absorbed RF power (P) dP/dH curve as a function of the applied magnetic field. The sample dimension for the FMR measurement was around 3 mm × 3 mm. The FMR spectra were registered by sweeping the applied field angle around the sample plane (SP) and sample normal (SN). In SP measurement, the magnetic component of the microwave field is always in the film plane, whereas the external magnetic field is rotated from the film plane towards the film normal. In SN measurement, the magnetic component of the microwave field is perpendicular to the film plane and the external magnetic field is rotated in the film plane for each spectrum. Representative FMR spectra of the YIG films in SP configuration were given in Fig. 3(c) for the applied field direction along with the film normal and in the film plane (Figs. 3(a) and 3(b)).

When the applied field was parallel to the film normal, the spectrum was at low field (red spectra), whereas it shifted to higher field (black spectra) when the applied field was parallel to the film plane, for all samples. This behavior refers that the easy axis of the magnetic anisotropy is perpendicular to the film plane. In SN geometry measurements, there was no any anisotropic behavior, which is not surprising. Thin films having PMA do not represent any anisotropic behavior in the film plane [18, 23, 36].

Further analysis on intrinsic magnetic properties of the system is performed by angular FMR measurements and numerical calculations. To reveal the micromagnetic parameters of YIG/Si (100) structure, the energy Hamiltonian presented in Eq.1 is employed and numerically solved.

$$E = \begin{Bmatrix} -\sum_{i=1}^{N} M_{eff} H[cos(\theta_H) cos(\theta_i) + sin(\theta_H) sin(\theta_i) cos(\varphi_H - \varphi_i)] \\ +\sum_{i=1}^{N} [K_{eff} cos^2(\theta_i) + K_{eff\_q} cos^4(\theta_i)] \end{Bmatrix}$$

(1)

The Hamiltonian consists of two energy terms used to represent the magnetic behavior of the systems [33, 37]. Here, (θ, θH) and (φ, φH) are, respectively, the polar and azimuth angles for magnetization vector M and external DC magnetic field vector H with respect to the film plane. External DC magnetic field is represented by the first term of the Hamiltonian, i.e., Zeeman energy. Effective magnetic anisotropy energy consists of the demagnetization energy, the interface energy and the first-order term of magnetocrystalline energy of the system. And, the last term represents the second-order magnetocrystalline energy. In Eq. (1), Meff, Keff, and Keff_q are the effective magnetization, effective magnetic anisotropy energy density, second order term of magnetocrystalline energy density, respectively. We scan the DC magnetic field from 0 to 1 T to determine the field corresponding to the maximum value of the dynamic susceptibility, which is called as the resonance field (Hres). Dynamic susceptibility spectra are recorded by using the Soohoo formulation for ferromagnetic resonance in multilayer thin films [38-40].

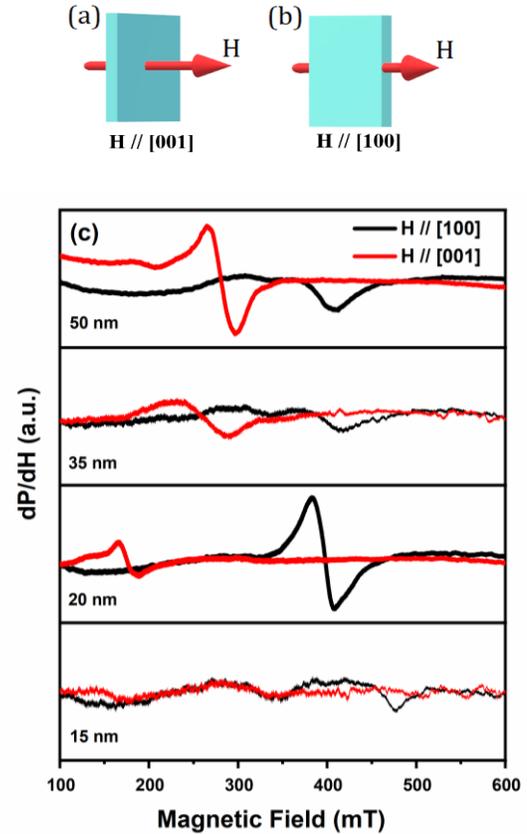

Fig. 3. FMR spectra of the YIG films in SP measurement geometry. (a) The applied field direction is along the film normal (H // [001] of Si substrate) and (b) along the film plane (H // [100] of Si substrate). (c) The black and red lines indicate the FMR spectrum when the magnetic field is parallel (H // [100]) and perpendicular (H // [001]) to the film plane.

The resonance fields are extracted from the recorded spectrum for SP geometry. By performing the aforementioned procedure for different angles of the magnetic field with respect to the film plane, we are able to reproduce the experimental data. All calculations were performed at room temperature. Meff, Keff, and Keff_q were obtained by the simulation model for all samples. Here, the total energy was minimized with the values of the magnetic parameters given in Table I. The angular dependence of the resonance field for different thicknesses in SP geometry is shown in Fig. 4.

Table I represents the result of the numerical calculations. The positive effective anisotropy energy density confirms that the easy axis is perpendicular to the film plane. The effective magnetization is lower than the bulk YIG value, which may be caused by possible crystal vacancies / deficiencies, inter diffusion between the substrate and the film, and Fe ion variation in the film [10]. In general, shape anisotropy is in the film plane. The increase in thickness strengthens the contribution of shape anisotropy to in-plane magnetic anisotropy, while the strain between the substrate-film tends to relax and, therefore, the effective perpendicular magnetic anisotropy reduces by increasing thickness as seen in Table I.

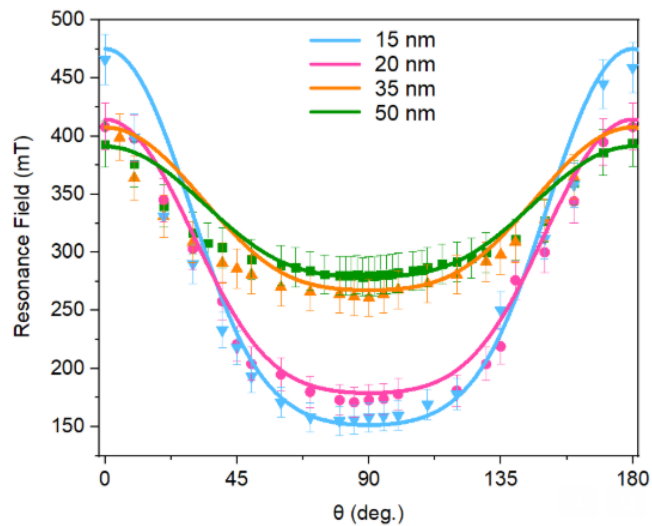

Fig. 4. A plot of angular variation of FMR resonance fields. Symbols and solid lines indicate the experimental and theoretical results; respectively.

Table I. Magnetic parameters obtained from the simulation for YIG thin films with PMA.

| Thickness(nm) | $K_{eff}$ (J/m$^3$) | $K_{eff\_q}$ (J/m$^3$) | $M_{eff}$ (kA/m) |
|---|---|---|---|
| 15 nm | 1773 | 320 | 105 |
| 20 nm | 1456 | 250 | 105 |
| 35 nm | 1260 | 150 | 105 |
| 50 nm | 1180 | 120 | 105 |

## IV. DISCUSSIONS

In this section, the structural and magnetic characterization of YIG films grown on Si (100) will be discussed. The crystallization of the films was analyzed by XRD measurements. Since we did not get the characteristic XRD peaks in as-grown films, an annealing process was required to generate the YIG phase [28]. After annealing at the temperature of 850 °C for 2 hours, we were able to observe (420) peak of the YIG from θ-2θ scan of the XRD measurement as seen in Fig. 1, which indicates the formation of the YIG phase. Some studies report the polycrystalline YIG film grown on quartz with three characteristic peaks [27, 32]. However, it seems that there is a preferential crystalline ordering or texturing in our films. When the film was annealed, the lattice of YIG locates on Si by making an angle of 26.6° between (400) plane of Si and (420) plane. There are two different crystallographic orientations / domain of the film lattice repeating each other on the substrate with respect to the symmetry axis of c, as shown in Fig. 5. The lattice mismatch is $(a_{film} - a_{substrate})/a_{substrate} = 1.94\%$, where "$a$" is the lattice constant. Since the lattice constant of YIG is larger than that of the substrate, a small compressive strain occurred at the interface, which can lead to a tetragonal distortion of the crystalline structure of the film.

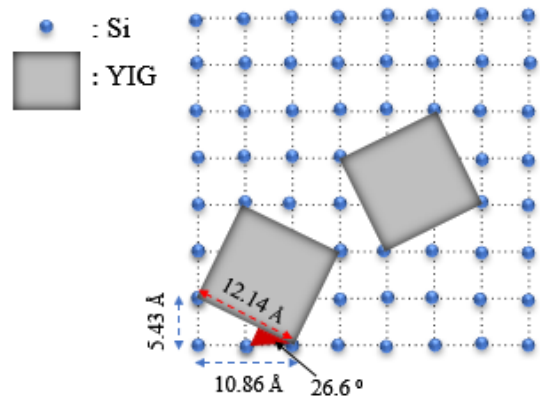

Fig. 5. A two-dimensional configuration of the lattice orientation of the film on Si substrate. Two preferential crystalline orderings/texturing were present in YIG after the annealing procedure.

The composition and electronic state of the Y-Fe-O elements on the surface of the YIG film were investigated by XPS analysis. The stoichiometry of the YIG film was determined by the percentage of O 1s, Fe 2p and Y 3p XPS peak areas shown in Figs. 2(b)-2(d) using relative sensitivity factors in CasaXPS software. The stoichiometry of our samples was found to be Y: 3.06, Fe: 5.17, O: 11.7 which is close to the expected one for Y:Fe:O of 3:5:12. The Fe percentage in our YIG film stoichiometry is 20.6 % which is similar to the ratio of 20% in bulk YIG as known from the literature [11]. Fig. 2d shows the core level Fe 2p spectra. Both $Fe^{3+}$ and $Fe^{2+}$ are present in the films [41]. 711.1 eV and 724.4 eV are the binding energy values for the $2p^{3/2}$ and $2p^{1/2}$ peaks of $Fe^{3+}$ and $Fe^{2+}$. Two peaks at 710.9 eV and 725.8 eV correspond to $Fe^{3+}$ $2p_{3/2}$ and $Fe^{3+}$ $2p_{1/2}$, and the binding energies at 708.86 eV and 724.16 eV refer to $Fe^{2+}$ $2p_{3/2}$ and $Fe^{2+}$ $2p_{1/2}$, respectively. The satellite structure of Fe $2p_{3/2}$ was located at 718.8 eV, binding energy higher than 710.9 eV. This shows that the Fe ions are in +3 valance states in the spectrum and located at tetrahedral sites of YIG lattice [28, 42, 43].

Representative FMR spectra of the samples are given in Fig. 3 for the external magnetic field parallel to the film normal and film plane. Resonance field is determined by taking the minimum value when applied along the easy axis.

The spectra clearly show PMA for all thicknesses. The FMR spectrum for H // [001] orientation of Si shifts towards higher field values as the thickness increases. In contrast, for H // [100] orientation of Si, the FMR spectrum shifts to lower values while increasing the thickness. This behavior indicates that the uniaxial perpendicular magnetic anisotropy decreases as the thickness increases. Magnetic properties of the films can be affected by many factors such as thickness, substrate, interfacial energy, and strain. The strain is compressive by 1.9% and tensile by 0.65% when the film is grown on Si (100) and GGG, respectively. In our case, the strain in the YIG films grown on the Si substrate is much greater than that grown on GGG. Due to the lattice mismatch between Si and YIG, available thickness with PMA, the value of magnetic anisotropies and FMR linewidth were different from studies using lattice-compatible substrates in the YIG thin film fabrication process. In this study, all the YIG thin films had a single uniform ferromagnetic resonance peak. In Fig.3, the shape and intensity of the FMR spectra vary depending on the thickness, crystalline quality, and magnetic homogeneity of films. Inhomogeneous broadening of the linewidth of the FMR spectra is due to the imperfections in film such as defects, roughness, symmetry breaking in surface and interface, oxygen vacancies, and inter ion diffusion between the layers [27]. The thinnest film has a wider and lower intensity FMR profile while the spectra gets clearer with the increase in thickness. Meanwhile, surface and interfacial strain effects show tendency to decrease and the increase in the amount of spins which interact with microwave field increases the FMR intensity. The reason for the FMR shape and intensity which do not vary in a systematic manner might be due to some uncontrollable parameters during deposition. However, it is observed that the out-of-plane magnetic anisotropy behavior of the films still exist in the pronounced thicknesses. The FMR linewidth was determined as the distance between the minimum and maximum point of the dP/dH curve, so called peak to peak linewidth. 20 nm YIG film has a linewidth of 230 Oe when the applied field is parallel to the sample plane and linewidth of 160 Oe when the applied field is perpendicular to the sample plane. The linewidths of the spectra are relatively larger compared to the reported value on GGG substrate [44]. However, there are similar linewidth values of that grown on quartz in the literature, as well [22, 27]. For example, 12 nm YIG film was reported to have an FMR linewidth of 250 Oe. For the film thickness range between 100 nm and 290 nm, linewidth values were between 340 Oe and 70 Oe. It is thought that defects due to the surface roughness and Fe iron deficiency may lead to magnon scatterings and increase of the FMR linewidth [11].

It is known that the magnetic easy axis in most of thin films are in the film plane due to the shape/dipolar anisotropy. Additional factor is necessary to overcome the shape anisotropy and switch the orientation of the easy axis from the film plane to the film normal. The crystalline or surface anisotropy or textured structure can trigger a perpendicular magnetic anisotropy [36]. Here, the lattice mismatch between Si and YIG thin film induced a compressive strain at the interface which led to a distortion of the lattice structure [4]. The compressive strain in the film plane results in an expansion along the c-axis, which switches the easy axis from the film plane to the film normal [36, 45]. In previous studies, PMA was realized in YIG films grown on different substrates in the thickness range of 10-20 nm [23, 32]. However, we achieved PMA up to 50 nm thickness as a result of texture and the lattice distortion of YIG.

In the literature, PMA in YIG films was observed in those grown on the buffer layer except GGG [23, 46]. This study indicates that PMA was attained successfully in YIG films on a non-garnet substrate without using any additional buffer layer or doping.

## V. CONCLUSION

YIG thin films with perpendicular magnetic anisotropy can pave the way for cutting-edge magnonic and spin-related technologies, i.e. for the fast response in microwave devices, logic devices, spin-transfer torque and magneto-optical device applications. Existence of the PMA in an insulator material is a rare magnetic phenomenon. In this work, we have achieved perpendicular magnetization in YIG thin films grown on Si substrate which is a common and base material of the present-day electronic industry. The effect of post annealing-temperature on the crystal structure and magnetic anisotropy was explored. XRD analysis revealed that the crystallization of YIG films improved after annealing. The compressive strain due to the lattice mismatch between Si and YIG led to a distortion in the YIG films, resulting in PMA in the thickness range of 15-50 nm. As far as our best knowledge, we report PMA in pure YIG thin films grown on Si substrate for the first time. We anticipate that perpendicular magnetized YIG thin films will allow the YIG magnetic insulator to be widely used in many areas.

### ACKNOWLEDGEMENTS

The authors are grateful to Dr. Ilhan Yavuz for his fruitful discussions on the results.